%% file: plos_latex_template.tex
\documentclass[10pt,letterpaper]{article}
\usepackage[top=0.85in,left=2.75in,footskip=0.75in]{geometry}

\usepackage{amsmath,amssymb}
\usepackage{bigints}
\usepackage[standard]{ntheorem}
\usepackage{changepage}
\usepackage{graphicx}

\usepackage{textcomp,marvosym}

\usepackage{cite}

\usepackage{nameref,hyperref}

\usepackage{microtype}
\DisableLigatures[f]{encoding = *, family = * }

\usepackage[table]{xcolor}

\usepackage{array}

\usepackage{subcaption}

\newcolumntype{+}{!{\vrule width 2pt}}

\newlength\savedwidth

\usepackage[aboveskip=1pt,labelfont=bf,labelsep=period,justification=raggedright,singlelinecheck=off]{caption}

\makeatletter
\renewcommand{\@biblabel}[1]{\quad#1.}
\makeatother

\usepackage{lastpage,fancyhdr,graphicx}
\usepackage{epstopdf}
\fancyheadoffset[L]{2.25in}
\fancyfootoffset[L]{2.25in}
\theoremstyle{break}

\newcommand{\ignore}[1]{ }

\newcommand{\cret}{\vspace*{\baselineskip}}

\newcommand{\beq}{\begin{equation}}
\newcommand{\eeq}{\end{equation}}

\definecolor{highlight}{rgb}{0,0,0} 

\usepackage{siunitx}
\usepackage{cite}
\usepackage{amsmath,amssymb,amsfonts}
\usepackage{algorithmic}
\usepackage{graphicx}
\usepackage{textcomp}
\usepackage{xcolor}
\usepackage{comment}
\usepackage{xspace}
\usepackage{wrapfig}
\usepackage{textcomp}
\usepackage{xspace}
\usepackage{compactbib}
\usepackage{array}
\newcolumntype{L}{>{\arraybackslash}m{4.5in}}
\usepackage{amsfonts}
\usepackage{amsmath}
\usepackage{amssymb}
\usepackage{amsmath}
\usepackage{algorithmic}
\usepackage{algorithm}
\usepackage[version=4]{mhchem}
\usepackage{caption}
\usepackage{subcaption}
\usepackage{xcolor,soul}

\begin{document}
\vspace*{0.2in}

\begin{flushleft}
   {\Large\textbf\newline{Neural network execution using nicked DNA and microfluidics}}
    \newline
   Arnav Solanki\textsuperscript{*}\textsuperscript{1},
   Zak Griffin\textsuperscript{*}\textsuperscript{2},
   Purab Ranjan Sutradhar\textsuperscript{2},
   Amlan Ganguly\textsuperscript{2},
	Marc Riedel\textsuperscript{1$\dag$}
	\\
	\bigskip
   \textsuperscript{*} These authors contributed equally. \\
	\textsuperscript{1} Department of Electrical and Computer Engineering, University of Minnesota Twin-Cities, Minneapolis, MN, USA  \\
     \textsuperscript{2} Department of Computer Engineering, Rochester Institute of Technology, Rochester, NY, USA  \\

	%
	%





    $\dag$ Email of corresponding author: mriedel@umn.edu
\end{flushleft}

\section{Abstract}

DNA has been discussed as a potential medium for data storage. Potentially it could be denser, could consume less energy, and could be more durable than conventional storage media such as hard drives, solid-state storage, and optical media. However, computing on data stored in DNA is a largely unexplored challenge. This paper proposes an integrated circuit (IC) based on microfluidics that can perform complex operations such as artificial neural network (ANN) computation on data stored in DNA. It computes entirely in the molecular domain without converting data to electrical form, making it a form of \emph{in-memory} computing on DNA. The computation is achieved by topologically modifying DNA strands through the use of enzymes called nickases. A novel scheme is proposed for representing data stochastically through the concentration of the DNA molecules that are nicked at specific sites. The paper provides details of the biochemical design, as well as the design, layout, and operation of the microfluidics device. Benchmarks are reported on the performance of neural network computation.



 
\section{Introduction}

\input{intro}

\input{multiplication}

\label{sec:multiplication}


\section{DNA-based Neural Engine}
\label{sec:dna-neural-engine}

\begin{figure}
\captionsetup[subfloat]{justification=centering}
\centering
\subfloat[]{
    \includegraphics[width=0.4\textwidth]{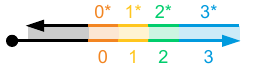}
    \label{fig:double_nick_1}
}\\
\subfloat[]{
    \includegraphics[width=0.4\textwidth]{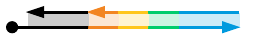}
    \label{fig:double_nick_2}
}\\
\subfloat[]{
    \includegraphics[width=0.4\textwidth]{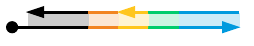}
    \label{fig:double_nick_3}
}\\
\subfloat[]{
    \includegraphics[width=0.4\textwidth]{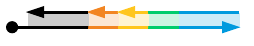}
    \label{fig:double_nick_4}
}
\caption{Storing data on DNA molecules using nicks. (a) The DNA template molecule consists of domains 0 to 4 (in color), with an additional unnamed domain (black) preceding them and a magnetic bead attached (on the left). 0*-4* denote the complementary top strand sequence for these domains. (b) The DNA molecule with a nick at nicking site $A$ between the black domain and 0*. (c) The DNA molecule with a nick at nicking site $B$ between the 0* and 1*. (d) The DNA with nicks on both nicking sites. Only this DNA molecule with two nicks represents data value 1; the other three configurations (a)-(c) correspond 0.}
\label{fig:double_nick}
\end{figure}

ANN computational workload consists primarily of matrix operations and activation functions. Among the matrix operations, matrix-matrix multiplication (GEMM) and matrix-vector multiplication (GEMV) make up almost the entirety of the workload which can be performed via repeated multiplications and accumulations (MAC).   
In the proposed DNA Neural Engine the process of performing a multiplication will take advantage of the stochastic representation of the operands. The input to a single neuron can be stochastically represented by the proportion of DNA strands nicked at a consistent site, compared to the total number of DNA strands in a solution (\textit{i.e.,} the concentration of specifically nicked DNA strands). In this paper, molecules with 2 nicks as shown in Fig. \ref{fig:double_nick} represent value 1, while all other molecule types correspond to 0. The relative concentration of doubly-nicked DNA molecules is the stochastic value stored in the solution.

The neuron weights, on the other hand, are represented by the concentration of enzymes in a droplet intended to create a second nick on the already-nicked DNA molecules.
To perform the stochastic multiplication for each neuron’s input-weight pair, the droplet with a concentration of enzymes, representing the weight value, is mixed with the droplet of the nicked DNA strands to create a second nick in the DNA strands. 
The second nicking site is required to be within around 18 base pairs of the first nick to allow a small fragment between the two nicked sites to be detached upon the introduction of probe strands. 
The product of the input and weight for this particular neuron is represented by the relative concentration of double-nicked strands compared to the total concentrations of DNA strands. 

It may be noted that at the beginning of the processing the inputs to the neural engine may also be set by this multiplication process where a solution of un-nicked DNA strands are nicked in a single site by the nickase enzymes whose concentrations are set to represent the input values thereby, creating an array of solutions with DNA strands with a single nick in concentrations representing the concentrations of the nickase and therefore the values of the inputs. Next, we describe the DNA-based neural engine hardware proposed in this work followed by the execution of the basic operations for an ANN.

\subsection{Neural Engine Architecture}

\begin{figure}[t]
  \centering
  \includegraphics[width=0.5\textwidth]{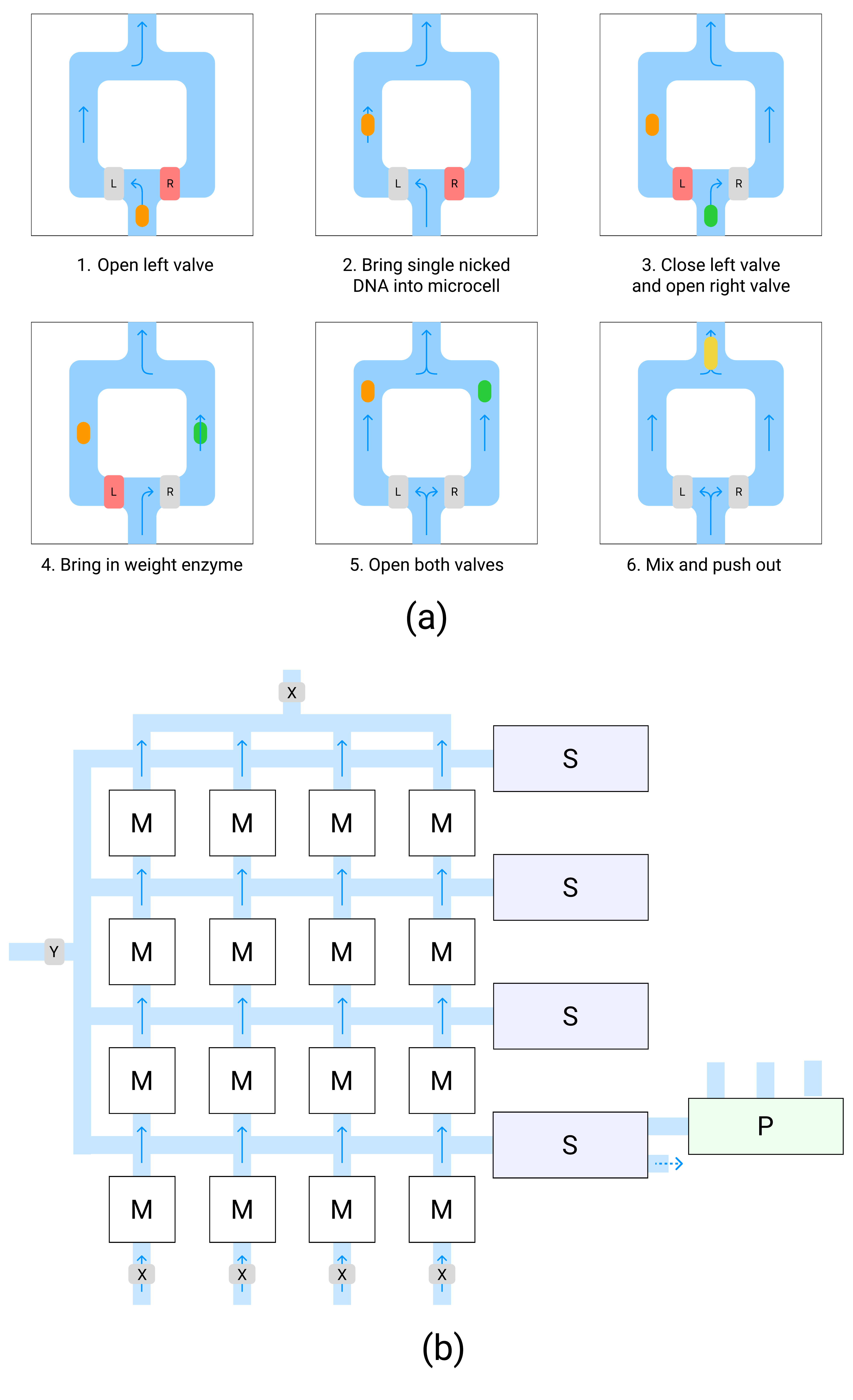}
  \caption{(a) Microcell operation sequence, and (b) Microcell assembly for Matrix Multiplications. The microfluidic channels are painted blue, with arrows showing flow direction induced by pressure differentiation. The gray and red boxes respectively represent Quake valves open and closed.}
  \label{fig:matrix}
\end{figure}

For the implementation of this process, we adopt a lab-on-chip (LoC) architecture. 
LoC emulates the electric signals in a digital chip with a set of controlled fluid channels, valves, and similar components. In our implementation, we will be using microfluidics where components are on the scale of 1-100$\mu$. Our system will operate using droplet-based microfluidics, meaning the fluid that holds data such as DNA or enzymes will move in small packages called droplets. The movement of droplets through the system will be controlled by creating pressure differentials. One critical component for controlling the flow of the microfluidic channels is the Quake valve which operates by running a pneumatic channel perpendicularly over a microfluidic channel. When the pneumatic channel is pressurized, it expands, closing the flow across the two sides of the microfluidic channel. To contain each stochastically nicked DNA droplet and merge these with weight enzymes, a small droplet storage container, which we will call a microcell, will be used as seen in Figure \ref{fig:matrix}(a).

\subsection{Microcell Function}
\label{microcell subsection}
Figure \ref{fig:matrix}(a) shows the sequence used to load and mix the two droplets holding the stochastically nicked DNA and weight enzymes. Throughout the loading, mixing, and release processes, there will be a constant pressure difference between the bottom and the top of the microcells shown in the figure, creating the upward flow into the next microcell. The steps, as demonstrated in Figure \ref{fig:matrix}(a), are described below:
\begin{enumerate}
    \item {The right valve R is closed, and the left valve L is kept open. This has the effect of routing the fluid through the left side of the microcell, leaving the fluid on the right-side static.}
    \item {The droplet of stochastically nicked DNA enters the microcell and continues until it is known to be at a predefined, timed distance along the left channel.}
    \item{The left valve is closed, and the right valve is opened, rerouting the fluid to flow along the right channel.}
    \item{The weight enzyme droplet is inserted into the microcell and continuously until it is known to be approximately the same distance along the right channel. It can be observed that the DNA droplet does not move since it is in static fluid.}
    \item{Both valves are opened, pushing both droplets simultaneously.}
    \item{The two droplets exit the microcell together, mixing them as the channels merge.}
\end{enumerate}

\subsection{Microcell Assembly}

The microcells will be arranged in a $k\times k$ formation, each capable of holding and mixing two droplets. These $k^2$ microcells are interconnected with a mesh of microfluidic channels, as shown in Figure \ref{fig:matrix}(b). In this figure, M, S, and P respectively represent the microcells, the merge modules, and the closing reaction pipelines. When delivering the nicked DNA droplets, all right valves are closed, and all left valves are open. The droplets are arranged at fixed distances so will travel across the microcells until each contains a single droplet. The weight enzyme droplets will similarly be inserted as in steps 3 and 4 of the microcell operation, with the exception that the left and right valve states are swapped this time. All left and right valves are then opened to perform steps 5 and 6 of the microcell operation shown previously in Figure \ref{fig:matrix}(a) and described in Section \ref{microcell subsection}.

\section{Implementation of ANN Operation in the Neural Engine}
\label{sec:ann-neural-engine}

Using the principles of stochastic computing with DNA nicking, we implement the operations involved in an ANN using the above microfluidic neural engine. 

\subsection{Execution of a Multiplication in a Neuron}
We demonstrate the execution of a single multiplication within a microcell by mixing two droplets containing our operands. The multiplicand is a concentration of $t$ DNA strands, nicked at a known site $A$ at a concentration $a$ (as shown in Fig. \ref{fig:double_nick}). The multiplier $b$ is represented by the concentration of nicking enzymes. The nicking enzymes are responsible for weakening the bonds holding the strands together so that after mixing and reacting, the strands nicked at both sites are our product, $a\times b$. The multiplier is a droplet of the weight enzyme with a concentration:

\begin{equation}
E = b \times t \times (1/k).
\end{equation}

\begin{figure}
\captionsetup[subfloat]{justification=centering}
\centering
\subfloat[]{
    \includegraphics[width=0.9\textwidth]{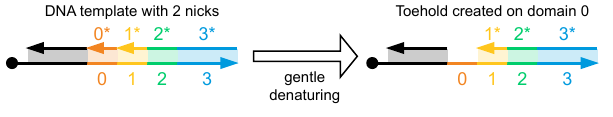}
    \label{fig:denature_probe_1}
}\\
\subfloat[]{
    \includegraphics[width=0.9\textwidth]{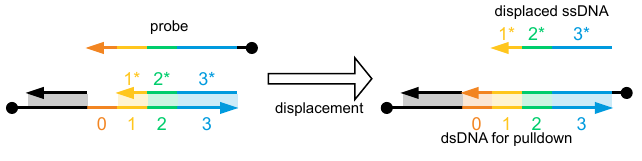}
    \label{fig:denature_probe_2}
}
\caption{Extracting ssDNA from dsDNA molecules using probe strands. (a) The DNA template molecule with two nicks at sites $A$ and $B$. After applying gentle heat, the ssDNA between the two nicks is selectively denatured to create a toehold at domain 0. (b) A probe strand is used to displace the ssDNA spanning domains 1 to 3 from the DNA molecule. The ssDNA is separated from all the other DNA molecules (i.e., the DNA and any excess probe strands) as the other molecules can all be pulled out.}
\label{fig:denature_probe}
\end{figure}

Here, $k$ represents the number of neurons present in the ANN layer, processed across $k$ microcells and the factor $1/k$ is a consequence of distributing the nicking enzymes over k microcells. To compensate for this $1/k$ operand, each of these nicking enzymes will be given enough time to react with $k$ DNA strands. This new nick will be at a second known site, $B$, nearby the first site $A$ as shown in Fig \ref{fig:double_nick_4}. This will result in $a \times t$ of the strands nicked at site $A$ and $b \times t$ of the strands nicked at site $B$. This means that the proportion of strands nicked at both sites will be the product of the two operands. A concentration of \textit{probe strands} are then introduced to displace the small ssDNA fragment from each of the aforementioned DNA product strands, as shown in Fig. \ref{fig:denature_probe}. The resulting proportion of free-floating ssDNA fragments with respect to the total DNA ($t$) strands represents the product, $ab$.

\subsection{Execution of Dot Product}

The above method for scalar multiplication can be used to compute the dot product for $k$ microcells, where each microcell contains the corresponding element of both input and weight vectors. Each of these $k$ microcells will undergo the multiplication as described, with the multiplier, $b$, being a unique weight enzyme concentration representing the weight values for each input pair. The products in each row of the microcell array as shown in Figure \ref{fig:matrix}(b) are then aggregated by mixing the droplets row-wise into one large combined droplet. This large combined droplet contains the sum of the number of fragments from each microcell which represents the dot product. Since the multiplicand in subsequent multiplications must be in the form of nicked DNA strands, this concentration of fragments must be transformed. Each fragment within the large droplet is mapped one-to-one to a nicking enzyme. This nicking enzyme is designed to nick at the primary site along a fresh, un-nicked DNA strand using a method known as strand displacement[]. 
The aforementioned method for dot product is implemented in the proposed microcell architecture using the following steps. 

\subsubsection{Droplet Merging}
\begin{figure}[t]

\includegraphics[width=1\textwidth]
{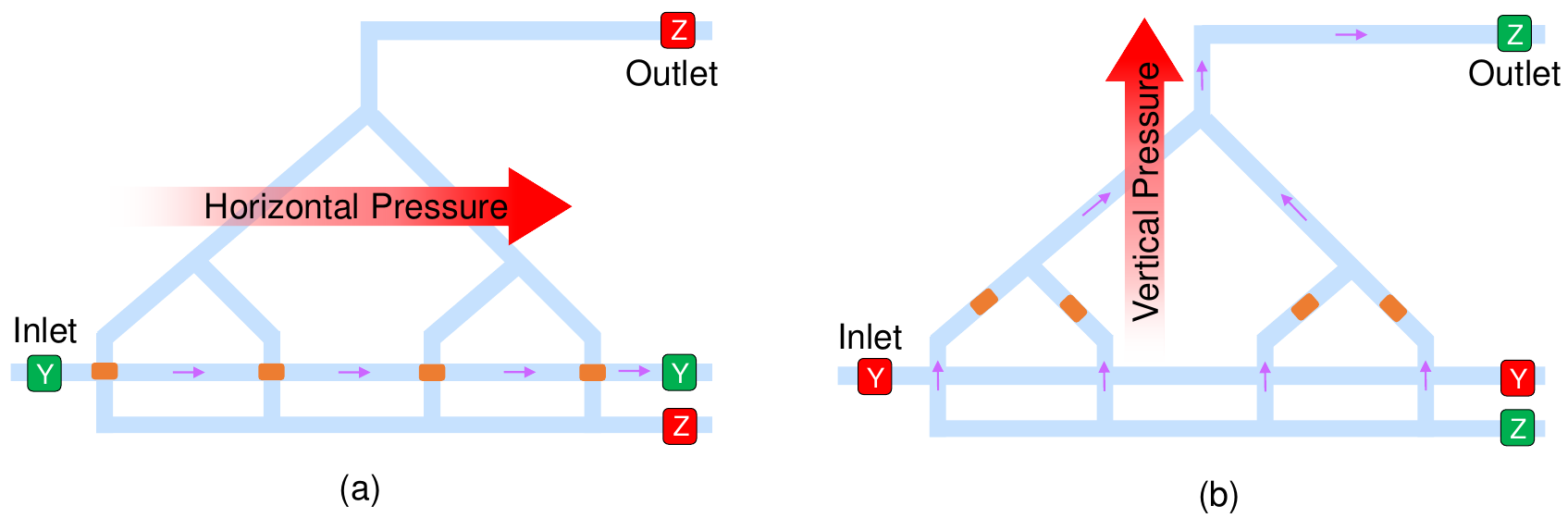}
\caption{The merging module, S.}
 \label{mergemodule}
\end{figure}

The droplet merging module, S shown in Figure \ref{fig:matrix}(b) adds the individual products of the elements of the two vectors to create the dot product. To compute the dot products as described, the mixed droplets from each microcell must be merged row-wise. Each droplet will exit the microcell, then take an immediate right turn, and remain on this horizontal path until entering the merging module, S. 
The two-step process is outlined as follows. Please refer to Figure \ref{mergemodule}.

\begin{enumerate}
\item{All droplets are merged into a single large droplet with the Y valves kept open (shown in green) and the Z valves closed (shown in red). This ensures a rightward flow and no vertical pressure difference. This is shown in Figure 10(a)}
\item{Next, the Y valves are closed (red), and the Z valves are opened (green), causing a pressure difference that forces each droplet upward through the merge channels. The construction of the merge channels is such that each droplet reaches the final merge point at the same time. This is shown in Figure 10(b)}
\end{enumerate}
Once each row of droplets has been mixed, they will go through the three-step closing reaction pipeline to apply the necessary transformations as discussed below.

\subsubsection{Reaction Pipeline}
The Reaction Pipeline module enables the implementation of an activation function in the DNA Neural Engine to the previously computed dot products. In addition to implementing the activation, it also transforms the nicked fragments into a singly-nicked DNA molecule to iteratively repeat the process to implement multiple ANN layers using the following steps. 

After merging all the droplets, the fraction of doubly nicked DNA molecules to all DNA molecules represents the dot product stored in the merged droplet as shown in Section \ref{sec:multiply}. By applying gentle heat to this droplet, toeholds are created on DNA molecules with two nicks due to partial denaturing.
The ssDNA next to this toehold can be displaced using probe strands as shown in Fig. \ref{fig:denature_probe}. Assuming complete displacement of these ssDNA molecules, the relative concentration (or to be even more precise, the relative number of molecules) of the ssDNA still represents the same fraction as the double-nicked DNA. Following this, we must apply an activation function on this ssDNA value to incorporate non-linear computations necessary in the neural networks.

Our approach utilizes a sharp sigmoid function with a user-defined transition point -- i.e., the activation function is a step function with the domain and range $[0,1]$, and the transition point can be set in the range $(0,1)$. This is achieved with the DNA seesaw gates presented by Qian and Winfree~\cite{Qian2011ScalingCascades}. This approach involves utilizing a basic DNA gate motif, which relies on a reversible strand-displacement reaction utilizing the concept of toehold exchange. The seesawing process allows for the exchange of DNA signals, with a pair of seesawing steps completing a catalytic cycle.  The reader is referred to \cite{Qian2011ScalingCascades} for further details.

\begin{figure}
\captionsetup[subfloat]{justification=centering}
\centering
\subfloat[]{
    \includegraphics[width=0.6\textwidth]{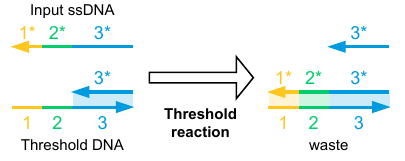}
    \label{fig:activation_1}
}\\
\subfloat[]{
    \includegraphics[width=0.9\textwidth]{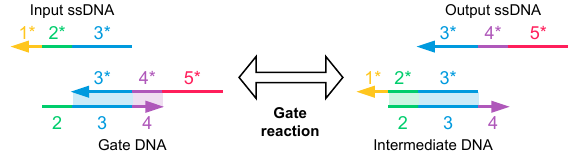}
    \label{fig:activation_2}
}\\
\subfloat[]{
    \includegraphics[width=0.7\textwidth]{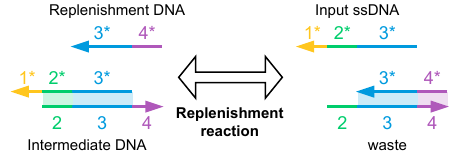}
    \label{fig:activation_3}
}\\
\subfloat[]{
    \includegraphics[width=0.7\textwidth]{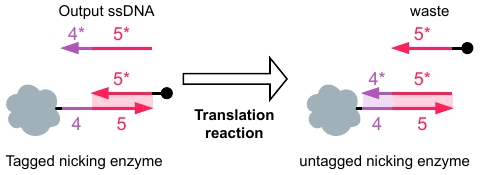}
    \label{fig:activation_4}
}
\caption{The set of reactions used to apply the activation function on ssDNA and generate an equivalent concentration of nicking enzyme. (a) The threshold reaction: the threshold molecule reacts with the input ssDNA to generate products that do not participate in any further reactions. (b) The gate reaction: the input ssDNA reacts with the seesaw gate molecule to create the output ssDNA and an intermediate molecule (c) The replenishment reaction: the replenishment strand reacts with the intermediate molecule to release more input ssDNA. This replenishes the concentration of input ssDNA and drives the production of more output ssDNA. (d) The translation reaction: the output ssDNA (domain 3* is not shown for clarity) reacts with the ``tagged" nicking enzyme (provided in excess) to produce an ``untagged" nicking enzyme. The concentration of untagged nicking enzyme is proportional to the concentration of the output ssDNA.}
\label{fig:activation}
\end{figure}

We use different DNA strands for thresholding and replenishing the output. The threshold molecule binds with the input ssDNA to generate waste (Fig \ref{fig:activation_1}), so the input ssDNA concentration must be larger than the threshold molecule concentration to preserve some residual amount of input ssDNA for the next stage. In the next stage, the gate reaction, the input ssDNA is used to generate output ssDNA (Fig \ref{fig:activation_2}). The replenishment strand in the (Fig \ref{fig:activation_3}) drives the gate reaction since it frees up more input ssDNA (Fig \ref{fig:activation_3}). That is, increasing the replenishment strand concentration maximizes the concentration of the output ssDNA \cite{Qian2011ScalingCascades}. 

With these DNA reactions, a gate can be designed that applies a threshold (in detail, the input ssDNA must be greater than the threshold DNA concentration) on the input ssDNA value, and then generates an output ssDNA value of 1 due to excess replenishment molecules. This allows us to implement a sigmoid activation function. If desired, the concentration of the replenishment molecules (Fig \ref{fig:activation_3}) can be limited to also apply an upper bound to the output ssDNA concentration.

With an activation function applied to the ssDNA concentration, we must now transform this value of DNA molecules to a value of nicking enzymes that can be used to trigger the next level of computation in the network. To achieve this, we will use a DNA strand displacement-based protein switch. First, we will conjugate the nicking enzyme with a DNA tag. This DNA tag will have one strand (called the \emph{major strand}) attached to the protein and contain a toehold, while the other strand (the \emph{minor strand}) will have a magnetic bead attached but will not connect with the protein directly. This is shown in Fig. \ref{fig:activation_4}. The DNA tag sequence will be constructed such that the toehold on the major strand will recruit the displaced DNA strands from the previous step, and the resulting strand-displacement reaction will entirely release the minor strand. The design of the protein-DNA tag allows individual displaced DNA strands to ``untag" nicking enzyme molecules. The remaining nicking enzymes (those that did not get to react with the DNA strands) will still be ``tagged" with magnetic beads and can be pulled out from the solution through the application of a magnetic field. After the pull-down process, the solution contains only untagged nicking enzymes at a specific concentration (this is discussed in detail below). This solution of nicking enzyme can now be used to nick site $A$ on a new droplet of DNA in the neuron downstream in the network.


\begin{enumerate}
\item Gentle heat is applied to the large, merged droplet. This allows denaturing of short DNA molecules and creates toeholds.
\item A droplet containing excess probe strands is mixed to release the input ssDNA fragments. The input ssDNA is separated from the remaining molecules through the application of a magnetic field.
\item A droplet containing the DNA seesaw gate, the threshold DNA (this amount is controlled by the user-defined sigmoid function), and the replenishment DNA (in excess) molecules is mixed with the ssDNA fragments. This applies a sigmoidal activation function on the ssDNA concentration.
\item The ssDNA strands are now mapped to a specific nicking enzyme concentration. For this, a drop containing an excess of the DNA-tagged nicking enzyme will be mixed with the ssDNA. After completion of the reaction, the drop will be subjected to a magnetic field to pull down the surplus nicking enzyme molecules. The resulting solution will contain the nicking enzyme with a concentration proportional to the particular concentration of the ssDNA strands after the activation function.
\item The droplet containing the nicking enzyme is now mixed with un-nicked DNA strands to prepare the inputs to the next layer of neurons in the ANN.


\end{enumerate}

After each stage of the reaction pipeline is completed, the merged droplets from each row now must be broken down into a collection of $k$ smaller droplets to be entered column-wise into the microcell array. This is accomplished using a droplet separator which functions by applying a pinching pressure at some regular interval to the channels carrying merged droplets~\cite{Berry19}. This results in a series of equally spaced droplets, which can then be placed back into the microcells column-wise.

\subsection{Layer-wise Execution of an ANN }

Using the $k \times k$ array of microcells and the S and P modules an entire layer of an ANN with $k$ neurons can be implemented.
In this array, each column implements a single neuron of the layer, and all the columns collectively form a single layer of the ANN. All microcells in the same column contain an equal nicked DNA concentration of double-stranded DNA molecules, $A_1-A_k$. The large droplet resulting from the output of each row's activation function is now divided back into $k$ originally sized droplets, which are then entered back into the microcell array column-wise, to repeat the computations for the next layer of the ANN, with the new inputs to each neuron held within the microcells.


\section{Results}
\label{sec:results}

In this work, we evaluate the proposed DNA Neural Engine while processing a simple ANN using the microfluidics-based DNA computing architecture in terms of latency of processing and area footprint of the device. 

The time for execution of a single layer, $t_\text{layer}$ can be modelled as follows:

\begin{equation}
t_\text{layer} = t_\text{transport} + t_\text{mult} + t_\text{merge} + t_\text{activation}.
\end{equation}

And, 
\begin{equation}
t_\text{activation} = t_\text{displacement} + t_\text{threshold} + t_\text{gate}  + t_\text{translation} + t_\text{nick}.
\end{equation}

Here: 
\begin{enumerate}
\item $t_\text{transport}$ is the time it takes for all droplets to travel throughout the microfluidic channels for all stages in the process. It is assumed that the time taken just for transportation is not the dominant bottleneck, and so it has been estimated to be around 2 minutes.
\item $t_\text{mult}$ is the time taken to perform a multiplication. This is the time taken for the second nicking of the strands, the second factor in the multiplication.

\item $ t_\text{merge}$ is the time taken to merge each of the small droplets per row into a single large droplet, the major step of the dot product summation.
\item $ t_\text{activation}$ can be broken up into several parts: displacement, inhibit, and nicking.
\item $ t_\text{displacement}$ is the time it takes to displace each of the ssDNA fragments from the doubly nicked strands
\item $ t_\text{threshold}$ is the time it takes for some input ssDNA strands to react with the threshold DNA. 

\item $ t_\text{gate}$ is the time it takes for the displacement of the output ssDNA alongside the replenishment reaction being used to drive the gate reaction.

\item $ t_\text{translation}$ is the time it takes for ``untagging" the right concentration of nicking enzyme and separating it.

\item $ t_\text{nick}$ is the time it takes for the untagged nicking enzyme to react with the fresh DNA strands for the resultant node value.
\end{enumerate}


The size of the proposed microfluidic device will scale quadratically with the number of neurons in a layer of the ANN, $k$ to support parallel execution of all neurons. This is because any layer with $k$ neurons requires an array of $k\times k$ microcells. As a pessimistic estimate, we assume each microcell will occupy an area equivalent of 6 channel widths of space in both length and breadth, given their structure with 2 microfluidic channel tracks in both horizontal and vertical directions as well an empty track for separation between the channels. Each track is assumed to be twice in width compared to the width of a channel to allow for manufacturability of the system. 

The following expression shows the area of a microcell array, $W$ with $k\times k$ microcells, where $c$ representing microfluidic channel width

\begin{equation}
W = s(c) = (6kc)^2.
\end{equation}

 A pessimistic channel width of $200\si{\micro\meter}$ yields a resulting expression for area of $(6\times 0.2 \times k)^2$ =$1.44k^2\si{\mm^2}$ for the array \cite{tan2004design}. 
 For an optimistic estimate, assuming a channel width of $35\si{\micro\meter}$, and a condensed microchamber estimate of $3\times 3$ channel widths per cell, we get an area estimate of $0.01k^2\si{\mm^2}$ for the microcell array \cite{tan2004design}. 
So depending on the manufacturing technology and fabrication node adopted, the parallelism of the device can be scaled up significantly to accommodate large hidden layers.

Table 1 shows the size and timing parameters of the microfluidic architecture \cite{tan2004design}. Here we assume that all neurons of a single layer of the ANN can be accommodated in the device simultaneously. Using these parameters we estimate the area requirements and delay for implementation of a simple ANN capable of classifying MNIST digits [refs]. In Table 2 we show the area and delay of the ANN for various device dimensions. The area estimate considers both a pessimistic and an optimistic dimension of the microfluidic channels and chambers from a fabrication perspective. We have considered multiple configurations (Config-1 to Config-4) corresponding to different device dimensions capable of accommodating varying numbers of microcells. These configurations offer a trade-off between device size and delay in ANN processing. In Config-1, we consider the number of microcells in the microfluidic system as 196 $\times$ 196 which is capable of accommodating an ANN layer with 196 neurons. Therefore, to accommodate the input layer for the ANN that receives the 28 $\times$ 28 MNIST frames the computations are serialized by a factor of 4 to compute the whole frame. Similarly, the other configurations require serialization by factors of 16, 49 and 196, respectively. Besides the input later, the designed ANN has a single hidden layer of 784 neurons and an output layer with 10 neurons. The hidden layer is serialized with the same factor as the input layer while the output layer did not need any serialization as it has only 10 neurons except for Config-4 where it was serialized by a factor of 3. Based on the required serialization factor and due to the limited number of microcells in a die the delay of executing a single layer is modified as follows, 

\begin{math}
t_{layer} = ((k_{layer} / k_{physical}) * (t_{transport} + t_{mult})) + t_{merge} + t_{activation},
\end{math}

where \begin{math}k_{layer}\end{math} and \begin{math}k_{physical}\end{math} are the number of neurons in an ANN layer and the number of neurons that can be computed simultaneously on the microfluidic die respectively. The Python model of the ANN was constrained to consider only positive inputs and weights and yielded an accuracy of 96\% in all the configurations as the computation model was not altered in any of them. 

We use a sigmoid activation function in all the layers, implemented with ``seesaw"' gates \cite{Qian2011ScalingCascades}, as discussed above. This enables signal amplification in the form of a sigmoid function -- precisely what we need. Again, the reader is referred to \cite{Qian2011ScalingCascades} for further details.

We assume that the partial results of the serialized computation can be stored in the DNA solution medium in an external reservoir array \cite{labonchip} that is communicating with the microfluidic ANN system through a microfluidic bus interface where the reservoirs are indexed and routed using the valve-system of the microfluidic system to the appropriate micro-chamber corresponding to the appropriate neuron. 

\begin{table}[b] \label{properties1}
  \vspace{-0.5em}
  \caption{Summary of the estimated system performance}
  \begin{center}
  \vspace{-0.5em}
  
   \begin{tabular}{|c|c|}
  
    \hline
    \textbf{Attribute} & \textbf{Value}\\     \hline
    Delay of single ANN layer (t$_{layer}$)    & 8.07 hrs\\
    \hline
    Channel Width (Optimistic) & $35\si{\micro\meter}$ \\ 
    \hline
    Channel Width (Pessimistic) & $200\si{\micro\meter}$ \\
    \hline
    Microcell Area (Optimistic) ($W_{min}$) & $0.01 mm^2$ \\ 
    \hline
    Microcell Area (Pessimistic) ($W_{max}$) & $1.44 mm^2$ \\ 
    \hline
\end{tabular}
\end{center}
\label{sim}

\vspace{-8mm}

\end{table}

\begin{table}[t] \label{properties-2}
  \scriptsize
  \vspace{-0.5em}
  \caption{Summary of the estimated system performance}
  \begin{center}
  \vspace{-0.5em}
  
   \begin{tabular}{|c|c|c|c|c|}
  
    \hline
    \textbf{Configuration}&\textbf{\# Microcells/}& \textbf{Microcell Array} & \textbf{Microcell Array} & \textbf{Execution}\\  &\textbf{Die}
    &\textbf{Area Pessimistic}& \textbf{Area Optimistic}& \textbf{ Time/Layer }\\ & &(cm$^2$) & (cm$^2$) & (hrs.) \\

    \hline
    Config-1 & $196 \times 196$ & 553.19 & 3.84 & 14.17  \\
    \hline
    Config-2 & $49 \times 49$ & 34.57 & 0.24 & 38.6 \\
    \hline
    Config-3 & $16 \times 16$ & 3.69 & 0.03 & 105.6 \\
    \hline
    Config-4 & $4 \times 4$ & 0.23 & 0.002 & 404.6 \\
    \hline
\end{tabular}
\end{center}

\vspace{-8mm}

\end{table}


Note that a configuration that minimizes the computational delay of the ANN for MNIST classification evaluated here would need a system with an array of $784\times 784$ microcells to accommodate the entire input layer simultaneously. However, that would make the die size unrealistic. Therefore, such a system could consist of multiple smaller microfluidic dies integrated on a microfluidic interposer substrate capable of communicating between the dies enabling a scalable solution \cite{Ganguly_int}. This system with $784\times 784$ microcells would reduce the delay per layer of the ANN to 8.07 hours.

A distinct advantage of using the DNA-based approach is that the variability of DNA as a computing medium adds an interesting new factor to ANN training. Slight variations in any reaction in the process could be used as a natural source of drift in training. Iterative feedback from executing the model could be used to correct the errors and further train the model indefinitely. This is not something reflected in traditional digital implementations without the artificial introduction of variation or noise between the models.

\input{conclusions}



\bibliography{references}

\end{document}

%% file: intro.tex
This paper presents a novel method for implementing mathematical operations in general, and artificial neural networks (ANNs) in particular, with molecular reactions on DNA in a microfluidic device. In what follows, we discuss the impetus to store data and perform computation with DNA. Then we outline the microfluidic technology that we will use for these tasks. 

\input{background}

\input{stochastic-logic}

\input{dna-strand-displacement}
\input{chemical-model}
\input{microfluidics}

\input{organization}

%% file: background.tex
\subsection{Background}

The fields of~\emph{molecular computing} and \emph{molecular storage} are based on the quixotic idea of creating molecular systems that perform computation or store data directly in molecular form. Everything consists of molecules, of course, so the terms generally mean computing and storage in~\emph{aqueous} environments, based on chemical or biochemical mechanisms. This is in contrast to conventional computers, in which computing is effected~\emph{electrically} and data is either stored \emph{electrically}, in terms of voltage, in solid-state storage devices; or \emph{magnetically}, in hard drives; or \emph{optically} on CDs and DVDs. Given the maturity of these conventional forms of computing and storage, why consider chemical or biochemical means? \\
\vspace{1ex}

\pagebreak
\noindent The motivation comes from distinct angles:
\begin{enumerate}

\item Molecules are very, very \emph{\bf small}, even compared to the remarkable densities in our modern electronic systems. For instance, DNA has the potential to store approximately 1,000 times more data per unit volume compared to solid-state drives. Small size also means that molecular computing can be \emph{localized}, so it can be performed in confined spaces, such as inside cells or on tiny sensors.

\item In principle, molecular computing could offer unprecedented \emph{\bf parallelism}, with billions of operations occurring simultaneously.

\item In principle, molecular computing could consume much less \emph{\bf energy} than our silicon systems, which always need a bulky battery or wired power source.

\item The use of naturally occurring molecules with enzymes results in a more \emph{\bf sustainable} computer design without the use of toxic and unethically sourced raw materials.

\item Finally, molecular computing could be deployed {\it\bf in situ} in our bodies or our environment. Here the goal is to perform sensing, computing, and actuating at a molecular level, with no interfacing at all with external electronics. The inherent biocompatibility of molecular computing components offers the possibility of seamless integration into biological systems. 
\end{enumerate}

\vspace{1ex}
\noindent{\bf DNA Storage}
\vspace{1ex}

The leading contender for a molecular storage medium is DNA. Ever since Watson and Crick first described the molecular structure of DNA, its information-bearing potential has been apparent to computer scientists. With each nucleotide in the sequence drawn from the four-valued alphabet of $\{A, T, C, G\}$, a molecule of DNA with $n$ nucleotides stores $4^n$ bits of data. Indeed, this information storage underpins life as we know it: all the instructions on how to build and operate a life form are stored in its DNA, honed over eons of evolutionary time.

In a highly influential Science paper in 2012, the renowned Harvard genomicist George Church made the case that we will eventually turn to DNA for information storage, based on the ultimate physical limits of materials~\cite{church12}. He delineated the theoretical storage \textbf{capacity} of DNA: 200 petabytes per gram; the read-write \textbf{speed}: less than 100 microseconds per bit; and, most importantly, the \textbf{energy}: as little as $10^{-19}$ joules per bit, which is orders of magnitude below the femtojoules/bit ($10^{-15}$ J/bit) barrier touted for other emerging technologies. Moreover, DNA is stable for decades, perhaps even millennia, as DNA extracted from the carcasses of woolly mammoths can attest. In principle, DNA could outperform all other types of media that have been studied or proposed.

\ignore{Church's capacity numbers are based on the physical dimensions of DNA. No one knows how to pack DNA in single-copy form at this density. His read-write speed estimates are based on molecular chemistry: the time it takes for a single letter of DNA to bind to the end of a strand. His energy estimates are based on the chemical energy required to form such bonds. An actual system has to \emph{move} and \emph{mix} chemicals to synthesize DNA, at scale. The space, time, and power to do this must be part of the equation.\footnote{Existing state-of-the-art DNA storage systems use liquid-handling robotics. The power consumption of these is on the order of \emph{hundreds} of watts, so \emph{hundreds} of joules/sec~\cite{ceze19}. This for a system that only synthesizes \emph{kilobytes}/sec. So, at present, the state-of-the-art is 18 orders of magnitude off of Church's theoretical limit when it comes to energy.} While such details might be dismissed as engineering challenges, they are formidable.}
Of course, no one has yet built a DNA storage system that comes close to beating existing media (magnetic, optical, or solid-state storage). The practical challenges are formidable.  
Fortunately, DNA technology is not exotic. Spurred by the biotech and pharma industries, the technology for both sequencing (\emph{reading}) and synthesizing (\emph{writing}) DNA has followed a Moore's law-like trajectory for the past 20 years. Sequencing 3 billion nucleotides in a human genome can be done for less than \$1,000. Synthesizing a megabyte of DNA data can be done in less than a day. Inspired no doubt by Church's first-principles thinking, but also motivated the trajectory of sequencing and synthesis technology, there has been a groundswell of interest in DNA storage. 
The leading approach is the synthesis of DNA based on phosphoramidite chemistry~\cite{ceze19}. However, many other creative ideas and novel technologies, ranging from nanopores~\cite{chen20} to DNA origami~\cite{dickinson21}, are being deployed.

\vspace{1ex}
\noindent{\bf DNA Computing}
\vspace{1ex}

Beginning with the seminal work of Adelman a quarter-century ago~\cite{adleman1994molecular}, DNA computing has promised the benefits of massive parallelism in operations. Operations are typically performed on the \emph{concentration} of DNA strands in solution. For instance, with DNA strand displacement cascades, single strands displace parts of double strands, releasing single strands that can then participate in further operations~\cite{yurke00, soloveichik10, Qian11-JRS}. The inputs and outputs are the concentration values of specific strands. 

\begin{figure}[t]
	\centering
	\includegraphics[width=4.0in]{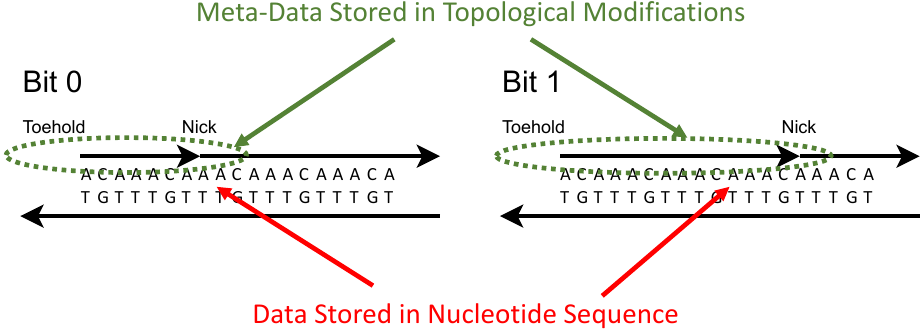}
   \caption{Data is stored in multiple dimensions. The sequence of nucleotides stores data in the form of the $A$'s, $C$'s, $T$'s, and $G$, with 2 bits per letter. Superimposed on this, we store data via topological modifications to the DNA, in the form of nicks and exposed toeholds. This data is \textbf{rewritable}, with techniques developed for DNA computation. }
	\label{fig:2D-DNA}
\end{figure}

It is fair to say that in the three decades since Adelman first proposed the idea, the practical impact of research on this topic has been modest. A practical DNA storage system, particularly one that is inherently programmable, changes this. Such storage opens up the possibility of ``in-memory'' computing, that is computing directly on the data stored in DNA~\cite{wang2019simd, chen21, Solanki23}. One performs such computation not on data stored not in the sequence of nucleotides, but rather by making topological modifications to the strands: breaks in the phosphodiester backbone of DNA that we call ``nicks'' and gaps in the backbone that we call ``toeholds.'' The nicking can be performed enzymatically with a system such as CRISPR/Cas9~\cite{jiang17, tabatabei20}. 

Note that the data that we operate on with this form of DNA computing is encoded in a different dimension than the data encoded in the sequence data of the DNA. The \textbf{underlying data} -- perhaps terabyte's worth of it -- is stored as the sequence of $A$'s, $C$'s, $T$'s, and $G$'s in synthesized strands. Superimposed on this, we store \textbf{metadata} via topological modifications. This is illustrated in Fig.~\ref{fig:2D-DNA}. This metadata is rewritable. Accordingly, it fits the paradigm of ``in-memory'' computing~\cite{ielmini18}. The computation is of SIMD form\footnote{SIMD is a computer engineering acronym for Single Instruction, Multiple Data~\cite{Flynn:1972}, a form of computation in which multiple processing elements perform the same operation on multiple data points simultaneously. It contrasts with the more general class of parallel computation called MIMD (Multiple Instructions, Multiple Data). Much of the modern progress in electronic computing power has come by scaling up SIMD computation with platforms such as graphical processing units (GPUs).} SIMD provides a means to transform stored data, perhaps large amounts of it, with a single parallel instruction.

%% file: stochastic-logic.tex
\subsection{Stochastic Logic}

The form of molecular computing that we present in this paper is predicated on a novel encoding of data. A link is made between the representation of random variables with a paradigm called \emph{stochastic logic} on the one hand, and the representation of variables in molecular systems as the concentration of molecular species, on the other. 

\begin{figure}[htpb]
	\centering
	\includegraphics[width=3.2in]{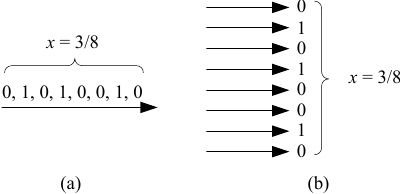}
	\vspace{1 ex}
	\caption{\small Stochastic representation: A random bitstream.  A value $x\in [0,1]$, in this case $3/8$, is represented as a bitstream. The probability that a randomly sampled bit in the stream is one is $x=3/8$; the probability that it is zero is $1-x=5/8$.}
	\label{Encoding}
\end{figure}

Stochastic logic is an active topic of research in digital design, with applications to emerging technologies~\cite{Gaines69, Qian10-TOC, Parhi15}. Computation is performed with familiar digital constructs, such as {\sc AND}, {\sc OR}, and {\sc NOT} gates. However, instead of having specific Boolean values of $0$ and $1$, the inputs are random bitstreams. A number $x$ $(0 \le x \le 1)$ corresponds to a sequence of random bits. Each bit has {\em probability} $x$ of being one and probability $1 - x$ of being zero, {\color{highlight} as illustrated in Figure~\ref{Encoding}.}
Computation is recast in terms of the probabilities observed in these streams.
Research in stochastic logic has demonstrated that many mathematical functions of interest can be computed with simple circuits built with logic gates~\cite{Qian10-TOC, Najafi17-TVLSI}. 
\\

\ignore{In a parallel model, these bits are assigned to wires in a bundle concurrently. In series the computation is probabilistic in \textit{time}, shown in~\ref{Encoding}(a); while in parallel the computation is probabilistic in {\em space}, shown in~\ref{Encoding}(b). All of our examples in this paper consist of serial stochastic computation.}

Consider basic logic gates. Given a stochastic input $x$, a {\sc NOT} gate implements the function
\begin{equation}
	\text{\sc NOT}(x) = 1 - x.
\end{equation}
This means that while an individual input of 1 results in an output of 0 for the {\sc NOT} gate (and vice versa), statistically, for a random bitstream that encodes the stochastic value $x$, the {\sc NOT} gate output is a new bitstream that encodes $1-x$.
The output of an {\sc AND} gate is $1$ only if all the inputs are simultaneously $1$. The probability of the output being $1$ is thus the probability of all the inputs being $1$. Therefore, an {\sc AND} gate implements the stochastic function:
\begin{equation}
	\text{\sc AND}(x, y) = x y,
\end{equation}
that is to say, multiplication. Probabilities, of course, are values between 0 and 1, inclusive. If we express them as rational numbers, given some positive integer $n$ as the denominator, we have fractions
\[
   x = \frac{a}{n}, \, y = \frac{b}{n}
\]
where $0 \leq a \leq n$ and $0 \leq b \leq n$. So an {\sc AND} gate computes \emph{a fraction of a fraction.} 

We can implement other logic functions. The output of an {\sc OR} gate is $0$ only if all the inputs are $0$. Therefore, an {\sc OR } gate implements the stochastic function:
\begin{equation}
	\text{\sc OR}(x, y) = 1 - (1-x)(1-y) = x + y - x y.
\end{equation}
The output of an {\sc XOR} gate is $1$ only if the two inputs $x, y$ are different. Therefore, an {\sc XOR} gate implements the stochastic function:
\begin{equation}
	\text{\sc XOR}(x,y) = (1-x)y+x(1-y)= x + y - 2 xy.
\end{equation}
The {\sc NAND}, {\sc NOR}, and {\sc XNOR} gates can be derived by composing the {\sc AND}, {\sc OR}, and {\sc XOR} gates each with a {\sc NOT} gate, respectively. Please refer to Table \ref{tab:stoch-gates} for a full list of the algebraic expressions of these gates. 
It is well known that any Boolean function can be expressed in terms of {\sc AND} and {\sc NOT} operations (or entirely in terms of {\sc NAND} operations). Accordingly, any function can be expressed as a nested sequence of multiplications and ~$1 - x$~ type operations. 
\cret

\begin{table}[htpb]
   \centering
   \caption{Stochastic Function Implemented by Basic Logic Gates}
   \label{tab:stoch-gates}
   \begin{tabular}{c|c|c}
      gate      & inputs & function           \\ \hline
      \sc{NOT}  & $x$    & $1 -x $            \\ \hline
      \sc{AND}  & $x, y$ & $x y$              \\ \hline
      \sc{OR}   & $x, y$ & $x + y - xy$       \\ \hline
      \sc{NAND} & $x, y$ & $1 - x y$          \\ \hline
      \sc{NOR}  & $x, y$ & $1 - x - y + xy$   \\ \hline
      \sc{XOR}  & $x, y$ & $x + y - 2 xy$     \\ \hline
      \sc{XNOR} & $x, y$ & $1 - x - y + 2 xy$
   \end{tabular}
\end{table}

There is a large body of literature on the topic of stochastic logic. We point to some of our prior work in this field. In~\cite{Qian10-EJC} we proved that any multivariate polynomial function with its domain and codomain in the unit interval $[0,1]$ can be implemented using stochastic logic. In~\cite{Qian10-TOC}, we provide an efficient and general synthesis procedure for stochastic logic, the first in the field. In~\cite{Qian11-TCAD}, we provided a method for transforming probabilities values with digital logic. Finally, in~\cite{Jenson16-ICCAD, Najafi19} we demonstrated how stochastic computation can be performed deterministically.

%% file: dna-strand-displacement.tex
\subsection{DNA Strand Displacement}

\label{dna_background}

DNA is generally present in double-stranded form ({\bf dsDNA}), in double-helix, with A's pairing with T's, and C's with G's. Without qualification, when we refer to ``DNA'' we mean double-stranded. However, for the operation we describe here, DNA in single-stranded form ({\bf ssDNA}) plays a role.

The molecular operation that we exploit in our system is called DNA strand displacement~\cite{yurke00, Seelig06}. It has been widely studied and deployed. Indeed, prior work has shown that such a system can emulate \textit{any} abstract set of chemical reactions. The reader is referred to Soloveichik et al. and Zhang et al. for further details ~\cite{soloveichik10,zhang2011dynamic}. Here we illustrate a simple, generic example. In Section~\ref{sec:ann-neural-engine}, we discuss how to map our models to such DNA strand-displacement systems.

We begin by first defining a few basic concepts. DNA strands are linear sequences of four different nucleotides $\{A, T, C, G\}$. A nucleotide can bind to another following \textit{Watson-Crick} base-pairing: A binds to T, C binds to G. A pair of single DNA strands will bind to each other, a process called \textit{hybridization}, if their sequences are complementary according to the base-pairing rule, that is to say, wherever there is an $A$ in one, there is a $T$ in the other, and vice versa; and whenever there is a $C$ in one, there is a $G$ in the other and vice-versa. The binding strength depends on the length of the complementary regions. Longer regions will bind strongly, smaller ones weakly. Reaction rates match binding strength: hybridization completes quickly if the complementary regions are long and slowly if they are short. If the complementary regions are very short, hybridization might not occur at all. (We acknowledge that, in this brief discussion, we are omitting many relevant details such as temperature, concentration, and the distribution of nucleotide types, i.e., the fraction of paired bases that are A-T versus C-G. All of these parameters must be accounted for in realistic simulation runs.)

\begin{figure}[!t]
	\centering
	\includegraphics[width=0.5\linewidth]{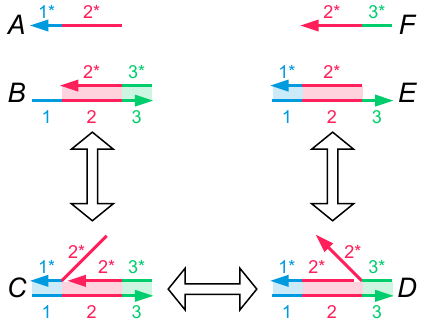}
	\caption{A set of DNA strand displacement reactions. Each DNA single strand is drawn as a continuous arrow, consisting of different colored domains numbered 1 through 3. DNA domains that are complementary to each other due to A--T, C--G binding are paired as $1$ and $1^*$.  The first reaction shows reactants A and B hybridizing together via the toehold at domain $1^*$ on molecule $B$. The second reaction depicts branch migration of the overhanging flap of DNA in molecule $C$, thereby resulting in the nick migrating from after domain $1$ to $2$. The third reaction shows how an overhanging strand of DNA can be peeled off of molecule $D$, thereby exposing a toehold at domain $3^*$ on molecule $E$ and releasing a freely floating strand $F$. All reactions are reversible. The only domains that are toeholds are $1^*$ and $3^*$.}
	\label{fig:displacement}
\end{figure}

Figure~\ref{fig:displacement} illustrates strand displacement with a set of reversible reactions. The entire reaction occurs as reactant molecules $A$ and $B$ form products $E$ and $F$, with each intermediate stage operating on molecules $C$ and $D$. In the figure, $A$ and $F$ are single strands of DNA, while $B$, $C$, $D$, and $E$ are double-stranded complexes. Each single-strand DNA molecule is divided, conceptually, into subsequences that we call \textbf{domains}, denoted as 1, 2, and 3 in the figure. The complementary sequences for these domains are $1^*, 2^*$ and $3^*$. (We will use this notation for complementarity throughout.) All distinct domains are assumed to be \textit{orthogonal} to each other, meaning that these domains do not hybridize.

\textbf{ Toeholds} are a specific kind of domain in a double-stranded DNA complex where a single strand is exposed. For instance, the molecule $B$ contains a toehold domain at $1^*$ in Figure~\ref{fig:displacement}. Toeholds are usually 6 to 10 nucleotides long, while the lengths of regular domains are typically 20 nucleotides. The exposed strand of a toehold domain can bind to the complementary domain from a longer ssDNA, and thus toeholds can trigger the binding and displacement of DNA strands. The small length of the toehold makes this hybridization reversible.
\ignore{A name consisting of a star means that the sequence of the domain is complementary to the original domain, for example, $A$ and $A^*$ are complementary to each other in DNA binding. The sequences of domains with different names are assumed to be orthogonal to each other, which is to say that they do not hybridize or react with strands other than their complementary domains.}

In the first reaction in Figure~\ref{fig:displacement}, the open toehold $1^*$ in molecule $B$ binds with domain $1$ from strand $A$. This forms the molecule $C$ where the duplicate $2$ domain section from molecule $A$ forms an overhanging flap. This reaction shows how a toehold triggers the binding of DNA strands. In molecule $C$, the overhanging flap can stick onto the complementary domain $2^*$, thus displacing the previously bound strand. This type of branch migration is shown in the second reaction, where the displacement of one flap to the other forms the molecule $D$. This reaction is reversible, and the molecules $C$ and $D$ exist in a dynamic equilibrium. The process of branch migration of the flap is essentially a random walk: at any time when part of the strand from molecule $A$ hybridizes with strand $B$, more of $A$ might bind and displace a part of $F$, or more of $F$ might bind and displace a part of $A$. Therefore, this reaction is reversible. The third reaction is the exact opposite of reaction 1 -- the new flap in molecule $D$ can peel off from the complex and thus create the single-strand molecule $F$ and leave a new double-stranded complex $E$. Molecule $E$ is similar to molecule $B$, but the toehold has migrated from $1^*$ to $3^*$. The reaction rate of this reaction depends on the length of the toehold $3^*$. If we reduce the length of the toehold, the rate of reaction 3 becomes so small that the reaction can be treated as a forward-only reaction. This bias in the direction of the reaction means that we can model the entire set of reactions as a single DNA strand displacement event, where reactants $A$ and $B$ react to produce $E$ and $F$. Note that the strand $F$ can now participate in further toehold-mediated reactions, allowing for cascading of such these DNA strand displacement systems.

%% file: chemical-model.tex
\subsection{Chemical Model}
\label{sec:model}

Recent research has shown how data can be encoded via {\em nicks} on DNA using gene-editing enzymes like CRISPR-Cas9 and PfAgo~\cite{Tabatabaei_Wang_Athreya_Enghiad_Hernandez_Leburton_Soloveichik_Zhao_Milenkovic_2019}. 
{\em Probabilistic switching} of concentration values has been demonstrated by the DNA computing community~\cite{Cherry_Qian_2018}. In previous work, we demonstrated how a concept from computer engineering called {\em stochastic logic} can be adapted to DNA computing~\cite{Salehi_Liu_Riedel_Parhi_2018}. In this paper,
we bring these disparate threads together: we demonstrate how to perform stochastic computation on
{\em fractionally-encoded} data stored on nicked DNA.

The conventional approach to storing data in DNA is to use a single species of strand to 
represent a value. It is either encoded as a binary value, where the presence of
the specific strand represents a 1 and its absence a 0~\cite{Seelig1585}; 
or as a non-integer value, encoded according to its concentration, called a {\em direct representation}~\cite{Wilhelm_Bruck_Qian_2018}.
In recent research, we have shown how a {\em fractional representation} can be used~\cite{Salehi_Liu_Riedel_Parhi_2018, Solanki23, Chen20ICASSP}. The idea is to use the concentration of two species of strand $X_0, X_1$ to represent a value $x$ with
$$x=\frac{X_1}{X_0+X_1}$$
where $x\in[0,1]$.
This encoding is related to the concept of {\em stochastic
logic} in which computation is performed on randomized bit streams, with values represented by the fraction of 1's versus 0's in the stream~\cite{Gaines1969},~\cite{Qian08-DAC},~\cite{Qian10-TOC}.

In this work, we store values according to nicking sites on double DNA strands. For a given site, we will have some strands nicked there, but others not. Let the overall concentration of the double strand equal $C_0$, and the concentration of strands nicked at the site equal $C_1$. The ratio of the concentration of strands nicked versus the overall concentration is
$$x=\frac{C_1}{C_0}$$
So this ratio is the relative concentration of the nicked strand at this site. We use it to represent a  variable $x \in [0,1]$.

Setting this ratio can be achieved by two possible methods. One is that we nick a site using a gene-editing guide that is not fully complementary to the nicking site. The degree of complementarity would control the rate of nicking and so set the relative concentration of strands that are nicked. A simpler method is to split the initial solution containing the strand into two samples; nick all the strands in one sample; and then mix the two samples with the desired ratio $x$. 

%% file: microfluidics.tex
\subsection{Microfluidics and Lab-on-Chip}

\begin{figure}[t]
  \centering
  \includegraphics[width=0.5\textwidth]{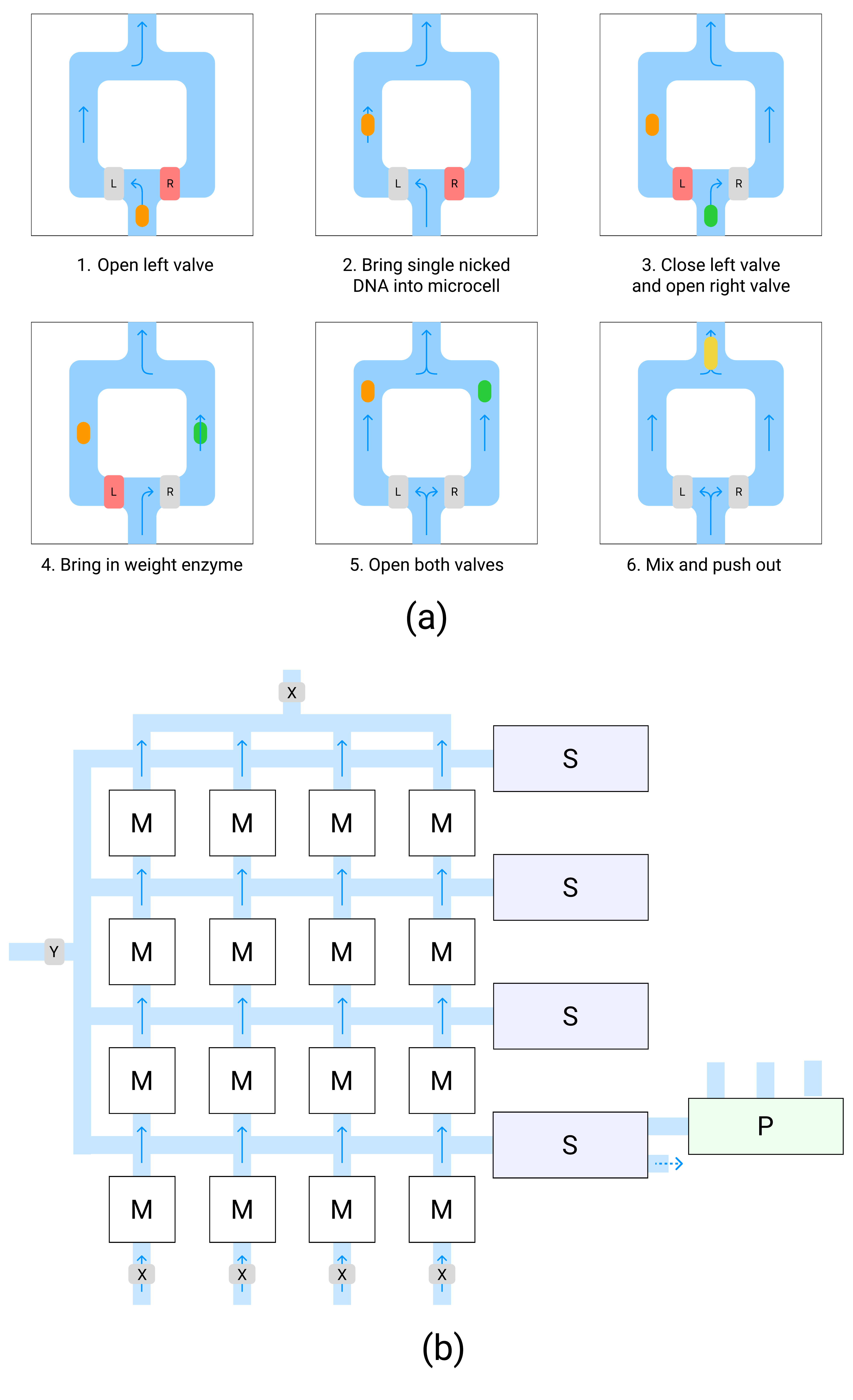}
  \caption{Microcell operation sequence. The microfluidic channels are painted blue, with arrows showing flow direction induced by pressure differentiation. The gray and red boxes respectively represent Quake valves open and closed.}
  \label{fig:microfluidics}
\end{figure}


Microfluidics is a rapidly developing discipline where small volumes of fluids are manipulated and transferred over channels whose dimensions range from one to hundreds of microns \cite{CONVERY2019}. Typically, such channels leverage principles of fluid dynamics enabling the modeling and design of systems where small volumes of fluids are moved to achieve a variety of purposes such as information and energy transfer. Due to their small form factors and need for very small amounts of fluids, this discipline is finding application in a variety of application domains such as cell sorting, DNA analysis, chemical synthesis and medical applications.

Utilizing the advances in microfluidics a practical device concept was envisioned as a Lab-on-Chip (LoC) \cite{Faraz2019}. A LoC is a device consisting of a network of microfluidic channels and microcells capable of transferring fluids to perform several functions such as chemical analysis, reactions, and sorting. Typical applications were in the area of medical sciences where small amounts of samples were needed to perform tests and diagnoses \cite{Faraz2019}. While the dominant application area of LoCs remains efficient medical diagnoses, advances in manufacturing capability using Integrated Circuit (IC) fabrication methodologies or 3D printing their applicability is expanding into sensing and processing more widely. In this paper, we envision an LoC device enabled by microfluidics to perform neural network computations using DNA molecules as the medium.

%% file: organization.tex
\subsection{Organization}

The rest of this paper is organized as follows. Section~\ref{sec:multiplication} describes how we implement our core operation, namely multiplication. We do so by computing a \emph{fraction} of a \emph{fraction} of concentration values. Section~\ref{sec:dna-neural-engine} presents the architecture of the microfluidic system that we use to implement computation on data stored in DNA. Section~\ref{sec:ann-neural-engine} discusses the implementation of an artificial neural network (ANN) using our microfluidic neural engine. Section~\ref{sec:results} simulations results of the ANN computation. Finally, Section~\ref{sec:conclusions} presents conclusions and discusses future work.

%% file: multiplication.tex
\section{Multiplication}
\label{sec:multiply}

The core component of our design is the multiplication operation, computed as a fraction of a fraction of a concentration value of nicked DNA. 

\subsection{Encoding Scheme}
\label{sec:encoding_nicking}
Nicking enzymes such as CRISPR-Cas9 can be used to effectively ``nick'' dsDNA at a particular site~\cite{jiang17, tabatabei20}. Since DNA is double-stranded, with strong base pairing between the A's and T's and the C's and G's, the molecule does not fall apart. Indeed, the nicking can be performed at multiple sites, and this process can be conducted independently. 

Suppose a molecule of DNA molecule with a particular nicking site labeled $A$ is in a solution. We separate the solution into two parts with a volume ratio $a$ to $1-a$ for some fraction $a$. Now site $A$ is nicked on all DNA molecules in the first solution, while the second solution is left untouched. These two solutions are mixed back to obtain a single solution. Some molecules in this solution are nicked, while others are not. The relative concentration of DNA molecules with a nick at site $A$ is $a$, while that of the molecules that are not nicked is $1-a$. Thus, any arbitrary fraction $a$ can be encoded in a solution of DNA molecules with a nicking site. In our framework, the stochastic value encoded at a particular site in DNA is the relative concentration (between 0 and 1) of DNA molecules with a nick at that site.

\subsection{Multiplying two values}
\label{sec:two_nicks}

\begin{figure}[t]
\centering
\includegraphics[width=0.65\textwidth]{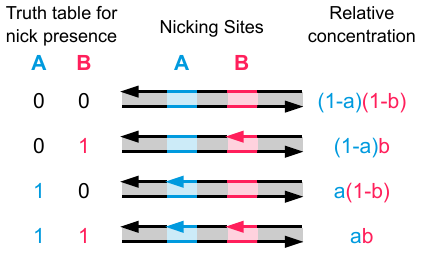}
\caption{Multiplying two values, $a$ and $b$, through nicking DNA. We start with a solution containing the DNA molecule shown on the top row. Fractional amount $a$  of these molecules are nicked at site $A$, and $b$ amount of all DNA molecules are nicked at site $B$. This results in a solution of 4 different possible DNA molecule types (as shown on each row). Assuming independent nicking on both sites, the concentration of each of these molecules is shown on the right. The molecule with nicks on both sites $A$ and $B$ has a concentration of $a\times b$, that is, the product of the two fractions.}
\label{fig:mult2}
\end{figure} 

Consider a DNA molecule with two unique nicking sites, $A$ and $B$. First, a stochastic value $a$ is encoded at site $A$, as was discussed in Section \ref{sec:encoding_nicking}. Now the single solution is again split into two parts, of volume ratio $b$ to $1-b$. All molecules are nicked at site $B$ in the first solution, while the second solution is again left untouched. Mixing these two solutions yields a solution containing DNA molecules that are either nicked at site $B$ or not. Thus, site $B$ now encodes the stochastic value $b$. Now both sites $A$ and $B$ are being used to independently store stochastic values $a$ and $b$. Since either site could be nicked or not nicked, there are 4 different possible molecules, as shown in Fig. \ref{fig:mult2}. 
Most significantly, the molecule containing two nicks, both at site $A$ and $B$, has a relative concentration of $a \times b$. That is the product of the two fractional values -- a fraction of a fraction. The concentrations of all other molecules are also listed in Fig.~\ref{fig:mult2}. Note that these values only hold if both sites are nicked independently.

Thus, our encoding approach not only allows us not only to store data but also to compute on it. This is ideal for computing a scalar multiplication in a neural network -- input data is initialized at site $A$ in a given solution, and then the scalar weight it is to be multiplied with is stored at site $B$. In this approach, it is necessary for sites $A$ and $B$ to be neighboring each other (i.e., no other nicking sites lie between them) to allow for readout.


\subsection{Reading Out}
\label{sec:readout}

\begin{figure}
\captionsetup[subfloat]{justification=centering}
\centering
\subfloat[]{
    \includegraphics[width=0.7\textwidth]{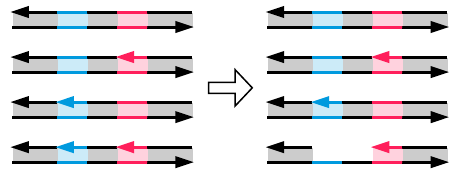}
    \label{fig:probe_1}
}\\
\subfloat[]{
    \includegraphics[width=0.72\textwidth]{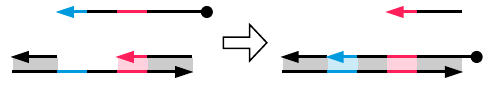}
    \label{fig:probe_2}
}
\caption{Reading out the multiplication results. (a) The DNA solution storing stochastic values $a$ and $b$ on sites $A$ and $B$ is gently heated. This creates a toehold only on the molecules with nicks on both sites, i.e., the $a\times b$ molecules. (b) A probe strand (the first reactant) can then bind with the newly exposed toehold and displace ssDNA (the first product). The concentration of this ssDNA stores the product $a \times b$.}
\label{fig:probeout}
\end{figure}

Having covered storing two stochastic values in a single solution, we now discuss multiplying these values. 

Assume a solution storing two stochastic values $a$ and $b$, as detailed in Section \ref{sec:two_nicks}. This solution is gently heated to initiate denaturing of DNA. That is, the DNA starts to break apart into two strands. By restricting the temperature, only short regions with low G-C content will fully denature, while longer strands remain bound. For our starting molecule, the short region between the nicking sites $A$ and $B$ will fully break apart into a single-stranded region. That is, a toehold will be formed between these two sites \cite{Hayward2108}. This toehold will only be formed on DNA molecules with nicks on both sites, so only $a\times b$ amount of molecules will have a toehold. Now a probe strand is supplied that will bind to the newly exposed toehold. This probe strand is used to displace the DNA strand adjacent to the toehold. The amount of single-stranded DNA (ssDNA) that is displaced through this process is again $a\times b$ the amount of the starting dsDNA. Thus, the product of two stochastic variables can be read out \textit{in vitro}. This procedure is shown in Fig. \ref{fig:probeout}. In Section \ref{sec:ann-neural-engine}, we discuss how these single strands can then participate in further strand-displacement operations.

It is important to cleanly separate the dsDNA molecules from the ssDNA extracted above. To achieve this, the dsDNA molecules and probe strands can have magnetic beads attached to them. When a magnetic field is applied to the solution, the dsDNA molecules and any excess probe strands can be pulled down from the solution, allowing the displaced ssDNA to be separated. These magnetic beads are shown in Fig. \ref{fig:denature_probe}.

%% file: conclusions.tex
\section{Conclusions}
\label{sec:conclusions}

Conventional silicon computing systems generally have centralized control with a CPU that can aggregate sensory data, execute arbitrarily complex analysis, and then actuate. For molecular applications, the actions of sensing, processing, and actuating must all be performed \emph{in situ}, in a decentralized way. Our goal in this paper was to devise molecular computing in which data processing occurs in the storage system itself using the natural properties of the molecules, with no need for readout and external electronic processing. \emph{In situ} molecular processing of data is critical from the standpoint of I/O: reading and writing data will always be a bottleneck for molecular systems. Computing ``in-memory'' is, therefore, a prerequisite.

We are collaborating with an industrial partner, Seagate, on the development of microfluidics technology for DNA storage. This technology will take many years to mature; however, when it does, techniques for computing \emph{on} the data that is stored in DNA will be needed. While conceptual in nature, this paper demonstrates how such computation could be performed.

In this paper presented a methodology for implementing complex operations, including ANN computation, on data stored in DNA. The paper weaves together two distinct strands: a conceptual representation of data, on the one hand, and the technology to compute with this representation, on the other hand. The representation is a fractional encoding on the concentration of nicked DNA strands. With this representation, we can compute a \emph{fraction} of a \emph{fraction} -- so the operation of multiplication -- borrowing ideas from stochastic logic. The ``read-out'' process is effected by releasing single strands via DNA toehold-mediated strand displacement. The technology is microfluidics. We described the microcell layout used in a pneumatic lab-on-a-chip (LOC) to control mixing. Mixing allows us to compute a fraction of a fraction of a concentration value. Based on this core operation, we presented a full architecture to implement neural computation.

There are a number of practical challenges. One of the concerns, ubiquitous with DNA strand displacement operations, is ``leakage'', that is to say errors in transforming concentrations. This occurs because we never have 100\% of DNA strands participating in designated reactions. Based upon the actual experimental results, we might have to mitigate leakage with error
correction methods or adopt so-called ``leakless'' designs~\cite{WangE12182}.

In future work, we will investigate ambitious applications of small molecule storage and computing. Our goal is to devise \emph{in situ} computing capabilities, where sensing, computing, and actuating occur at the molecular level, with no interfacing at all with external electronics. The applications include: 
\begin{itemize}
   \item {\bf Image processing and classification}:  We will implement a full-scale  molecular image classifier using neural network algorithms. Performing the requisite image processing {\it in situ}, in molecular form, eliminates data transfer bottlenecks. We will quantify the accuracy of image processing in terms of the \emph{signal-to-noise} ratio and the \emph{structural similarity index}.
   \itemsep0em

   \item {\bf Machine learning}: We will explore a common data representation for integrating sensing, computing, and actuation \emph{in situ}: hyperdimensional random vectors. Data is represented by long random vectors of integer or Boolean values. We will deploy this paradigm for machine learning, exploiting the randomness of molecular mixtures for encoding, which can naturally map to large vector representations.
\end{itemize}